\documentclass[twocolumn,showpacs,preprintnumbers,amsmath,amssymb]{revtex4}

\usepackage{graphicx}
\usepackage{dcolumn}
\usepackage{bm}

\newcommand{\be}{\begin{equation}}
\newcommand{\en}{\end{equation}}
\newcommand{\bea}{\begin{eqnarray}}
\newcommand{\ena}{\end{eqnarray}}

\begin{document}


\title{Minimal Brownian Ratchet: An Exactly Solvable Model}

\author{Youngki~Lee$^{1,2}$, Andrew Allison$^3$, Derek~Abbott$^3$, and
H.~Eugene Stanley$^2$}

\affiliation{
$^1$Yanbian University of Science \& Technology,
Beishan St. Yanji, Jilin, 133000, China \\
$^2$Center for Polymer Studies and Department of Physics,
Boston University, Boston, MA 02215 \\
$^3$Centre for Biomedical Engineering (CBME) and School of
Electrical and Electronic Engineering, The University of Adelaide,
SA 5005, Australia
}

\date{Last modified: June 19, 2003.  Printed: \today}

\begin{abstract}

We develop an analytically solvable three-state discrete-time minimal
Brownian ratchet (MBR), where the transition probabilities between
states are asymmetric. By solving the master equations we obtain the
steady-state probabilities. Generally the steady-state solution does
not display detailed balance, giving rise to an induced directional
motion in the MBR. For a reduced two-dimensional parameter space we
find the null-curve on which the net current vanishes and detailed
balance holds. A system on this curve is said to be balanced. On the
null-curve, an additional source of external random noise is
introduced to show that a directional motion can be induced under the
zero overall driving force. 

\end{abstract}

\maketitle




The Brownian ratchet and pawl system was first correctly explained by
Smoluchowski~\cite{smoluchowski1912} and later revisited by
Feynman~\cite{feynman1963} -- this has inspired much activity in the
area of Brownian ratchets, despite flaws in Feynman's analysis of the
thermal efficiency of the ratchet engine~\cite{parrondo1996} and
detailed balance~\cite{abbott2000}.

Interest has revived because molecular motors~\cite{astumian1994} have
been described in terms of Brownian
ratchet~\cite{doering1995,magnasco1993} models. Another area of
interest has been in Parrondo's paradox~\cite{harmer1999a} where losing
strategies cooperate to win. This can be illustrated in terms of games
that lose when played individually, but win when alternated -- this
has been shown to be a discrete-time Brownian ratchet~\cite{harmer1999b},
otherwise known as a `Parrondian game'. Parrondo's games have significantly
sparked recent interest in the areas of lattice gas automata~\cite{meyer02},
spin models~\cite{moraal00}, random walks and
diffusions~\cite{key02,pyke02,cleuren02}, biogenesis~\cite{davies01},
molecular transport~\cite{kinderlehrer02,heath02}, noise induced
patterns~\cite{buceta02}, stochastic
control~\cite{allison01a,kocarev02}, stochastic
resonance~\cite{allison01b} and quantum game
theory~\cite{flitney02,lee02}. Recently, Reimann~\cite{reimann02} has performed
an extensive review of the ratchet field.

Jarzynski {\it et~al.}~\cite{jarzynski1999} developed an exactly
solvable Brownian ratchet that can be operated as a heating system or
refrigerator, depending on the parameters between two heat reservoirs
of different temperatures. However this is treated as a six state
system and solution is via matrix inversion of coupled linear
equations. The derivation is somewhat complex, so the physical picture
and key ingredients of the observed properties are obscured.

Westerhoff {\it et~al.}~\cite{westerhoff1986} have analyzed enzyme
transport using a four-state model. In this paper, for the first time,
we develop a three-state discrete-time Brownian ratchet model that can
be solved analytically. We call it the {\it Minimal Brownian Ratchet}
(MBR)~\cite{footnote_01}. By setting up and solving the steady-state
solution of the corresponding master equations we obtain the
probability current and the null-surface, in the parameter space, of
the noisy and noise-free MBR. The obtained solution does not show any
critical behavior and can be suitably explained in terms of
non-singular behaviors.

\begin{figure}
   \resizebox{84.67mm}{!}{\includegraphics*{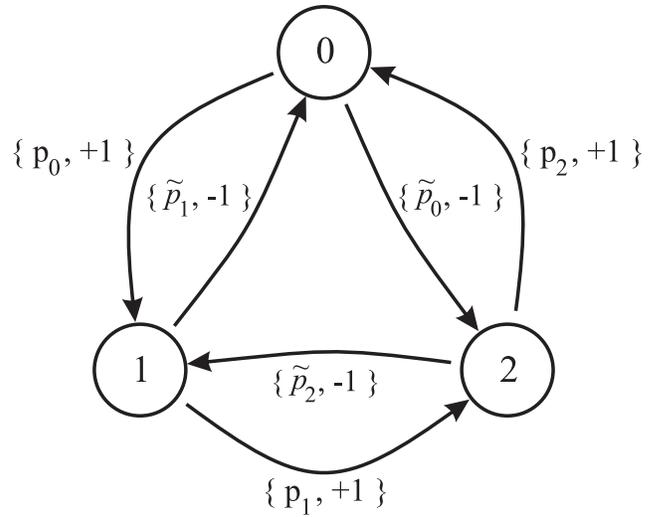}}
   \caption {State-transition diagram of a 3-state discrete-time Brownian
   ratchet with asymmetric transition probabilities $p_0$, $p_1$ and
   $p_2$ in the positive direction (counter-clockwise) and $(1-p_0)$,
   $(1-p_1)$ and $(1-p_2)$ in the negative direction (clockwise).  Each
   transition has two numbers associated with it, $\left\{ p_k , R_k \right\}$.
   The first number in the brackets, $p_k$, is the
   conditional probability of that transition (given the initial
   state). The second number, $R_k$, is the reward associated with that
   transition. Note that we have a skip-free process, which means the reward structure
   is $+1$ for `winning' transitions and $-1$ for `losing' transitions.}
   \label{fig:fig1}
\end{figure}

The minimal ingredients of a Brownian ratchet are an asymmetric potential and
random noise. In Fig.~\ref{fig:fig1} we show the state diagram of the
MBR. The MBR has three states, $\{ S_0,S_1,S_2 \}$, where the
transition probabilities between states are asymmetric. The transition
probability that a random walker in state $S_k$ steps in the positive
direction is $p_k$. The probability of a shift in the negative
direction is $\tilde{p_k}$. This is true for $k \in \{0,1,2 \}$. We
define the positive direction as counterclockwise.  The condition of
normalization, $p_k+\tilde{p_k}=1$, is automatically enforced by our
choice of symbols. These ingredients comprise a three-state random walk model with
generalized asymmetric potential and we call it a noise-free MBR.

It is straight forward to set up the following difference equations for
the probability distributions of the noise-free MBR model:
\be
  P_k(t+1)= P_{k+1}(t) \tilde{p}_{k+1} + P_{k+2}(t) p_{k+2}
  \label{eq:master}
\en
for all cyclic (modulo-3) state indices $k$.
$P_k(t)$ is the probability for the random walker at time $t$ to
be on the state of $S_k$. This can be written in matrix form as ${\bf
P }_{t+1} = {\bf P }_{t} B $, where ${\bf P}_{t}$ is the time varying
probability (row) vector at time $t$ and $B$ is the transition
probability matrix. We can write:
\be
  \left[B_{i,j} \right] =
  \left[
  \begin{array}{ccc}
  0            & p_0           & \tilde p_0  \nonumber \\
  \tilde{p_1}  & 0             & p_1                   \\
  p_2          & \tilde{p_2}   & 0           \nonumber
  \end{array}
  \right].
  \label{eq:transition_matrix}
\en
The steady-state probability, after a sufficiently long time,
$\lim_{t\rightarrow \infty} {\bf P_t} = {\bf P}$ is simply given as
\be
  {\bf P} = {\bf P} B
  \label{eq:sss}
\en
which is a characteristic value problem.
A partial probability current, $I$, can be defined as
\be
  I = P_kp_k - P_{k+1} \tilde{p}_{k+1}.
  \label{eq:current-definition}
\en
If $I=0$ there is no net current and detailed balance
\cite{onsager_1931} is satisfied, otherwise there exists a net current
and the system will assume a non-equilibrium steady-state.

Solving the Eq.~\ref{eq:sss} together with the normalization condition,
$   P_0+P_1+P_2=1$ ,
is again straightforward. Using the standard methods for
characteristic value or eigenvalue problems, we obtain
\be
 P_k = (\tilde{p}_{k+1} +p_{k+1}p_{k+2})/D
  \label{eq:steady-state-solutions}
\en
for all $k$. The denominator $D$ is given as
\be
  D = 2 + p_0p_1p_2 + \tilde{p_0} \tilde{p_1} \tilde{p_2}.
\en
These expressions are consistent with the results of
Pearce~\cite{pearce2000}. It is easy to check that they are the
solution to Eq.~\ref{eq:sss} by direct substitution.

We can substitute the results from Eqs.~\ref{eq:steady-state-solutions}
into Eq.~\ref{eq:current-definition} to solve for the net current,
$I$,
\be
  I = (p_0 p_1 p_2  - \tilde{p_0}\tilde{p_1} \tilde{p_2}) /D.
  \label{eq:noise-free-current}
\en
The condition for detailed balance $I=0$  is then
\be
  p_0 p_1 p_2 = \tilde{p_0} \tilde{p_1}\tilde{p_2},
  \label{eq:detailed-balance}
\en
which is the equation of a two-dimensional surface in the three
dimensional parameter space, $\{ p_0 , p_1 , p_2 \}$.  Note that the
Eq.~\ref{eq:noise-free-current} is independent of state index $k$ as
is required by its definition given in
Eq.~\ref{eq:current-definition}.

The second part of the MBR is to introduce  additional random noise
to the system.  To the noise-free MBR, we add more noise, controlled by the
parameter $\gamma$, to the MBR as follows.  With a probability of
$\tilde{\gamma}$, a random walker follows the dynamic rule of the
noise-free MBR otherwise, with the probability of $ \gamma~\left(
=1-\tilde{\gamma} \right)$, the walker randomly takes a right or left
step with the equal probability of a half. For $\gamma = 0$ the model
is exactly same with the noise-free MBR.  In the other limit, for
$\gamma = 1$, the randomizing process dominates and the system reduces
to a simple unbiased random walk where the net current remains zero.
It is important to note that $\gamma$ {\it influences} the level of
noise in the ratchet but is not {\it identical} with the noise itself.
We refer to $\gamma$ as a `noise parameter'. With this modification
the transition probability matrix $B$ changes as
\be
  \left[B_{i,j} \right] =
  \left[
  \begin{array}{ccc}
  0 & \tilde{\gamma}p_0 + \gamma /2 & \tilde{\gamma} \tilde{p_0} + \gamma /2   \nonumber \\
  \tilde{\gamma}\tilde{p_1} + \gamma /2 & 0 &  \tilde{\gamma} p_1 +\gamma /2             \\
  \tilde{\gamma}p_2 + \gamma /2 & \tilde{\gamma}\tilde{p_2} + \gamma /2   & 0  \nonumber
  \end{array}
  \right].
  \label{eq:transition-mbr}
\en

From the transition matrix, we know that adding the random noise with
parameter $\gamma$ effectively changes the existing parameters as
\be
  p_k \rightarrow \tilde{\gamma}p_k +\gamma /2
  \label{eq:parameter-change}
\en
and same holds for the $\tilde{p_k}$'s. The steady state solution and net
current for noisy MBR can be obtained by exchanging all the
$p_k$'s in Eqs.~\ref{eq:steady-state-solutions} and
\ref{eq:noise-free-current} according to
Eq.~\ref{eq:parameter-change}. The expression for current is given as
\be
  I_\gamma =[ \tilde{\gamma}^3A_- + \tilde{\gamma}(\gamma/2)(1-\gamma/2)B]/D_\gamma
\en
where
\bea
  D_\gamma &=& 2 + \tilde{\gamma}^2A_+ + \tilde{\gamma}(\gamma/2)(1+\gamma/2) \nonumber  \\
  A_{\pm}&=&p_0p_1p_2 \pm \tilde{p_0}\tilde{p_1}\tilde{p_2}                   \\
  B&=&p_0+p_1+p_2-\tilde{p_0}-\tilde{p_1}-\tilde{p_2}.                        \nonumber
\ena

It is possible to further restrict the choices of $\{p_0,p_1,p_2\}$
without losing the important properties of the ratchet. Parrondo's
original definition imposed the further constraints $p_0=q$ and
$p_1=p_2=p$.  This reduced the parameter space to a two dimensional
space with parameters $\{p,q\}$.  In two dimensional $\{p,q\}$
parameter space, the condition of detailed balance, i.e.,
Eq.~\ref{eq:detailed-balance}, gives the equation for a curve that we
call the null-curve:
\be
  q = {\tilde{p}^2 \over p^2 +\tilde{p}^2} = {(1-p)^2 \over p^2 + (1-p)^2}.
  \label{eq:null-curve}
\en
The null-curve is a special case of the more general
null-surface or null-hypersurface, in higher dimensions.
\begin{figure}
  \resizebox{84.67mm}{!}{\includegraphics*{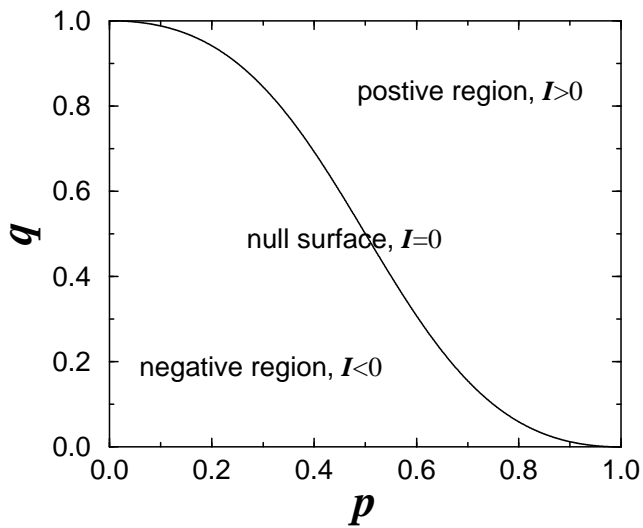}}
  \caption
  {The null-surface of a 3-state discrete-time Brownian ratchet.  On the
  null-surface, $q = (1-p)^2/(p^2+(1-p)^2)$, the current vanishes.
  Above the curve, the system has positive net current. Below the
  curve, the system has negative net current.}
  \label{fig:fig2}
\end{figure}
Fig.~\ref{fig:fig2} shows the `positive' and `negative' net current
regions of the noise-free MBR. Note that as expected from the symmetry
of the system the curve is invariant under the transformations $q
\rightarrow (1-q)$ and $p \rightarrow (1-p)$. This also apparent from
a consideration of Eq.~\ref{eq:detailed-balance}.

On the null-surface, we introduce additional random noise to the system
by controlling the value of $\gamma$. For $\gamma = 0$ the model is
exactly same as the noise-free MBR and the net current remains zero
since we are on the null-surface. In the other limit, for $\gamma =
1$, the randomizing process dominates and the system reduces to a
simple unbiased random walk where the net current is also zero.
However, counter-intuitively, for $0<\gamma<1$ non-zero current is
induced by introducing random noise controlled by $\gamma$.

\begin{figure}
  \resizebox{84.67mm}{!}{\includegraphics*{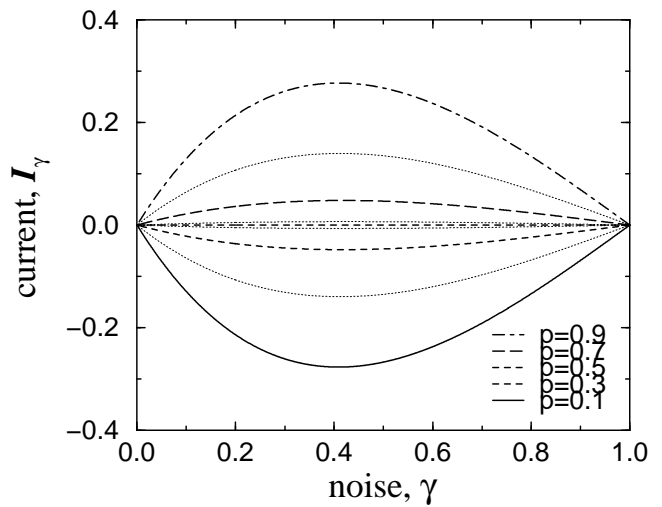}}
  \caption
  {The probability current, $I_\gamma$, versus noise parameter,
  $\gamma$, on the null-surface. For values of $p \neq 0.5$,
  additional noise induces a net current that increases, in
  magnitude, with increasing $\gamma$ and then decreases, in
  magnitude, to zero after $\gamma$ exceeds an optimum value. The
  bottom curve corresponds to $p=0.1$. All the other curves represent
  increments of $\Delta p = 0.1$. The middle curve corresponds to $p
  = 0.5$. The top curve corresponds to $p = 0.9$. Parrondo's original
  games had $p = 0.75$.}
  \label{fig:fig3}
\end{figure}

In Fig.~\ref{fig:fig3} we show the current versus noise parameter,
$\gamma$, for different values of parameters $p$ and
$q=\tilde{p}^2/(p^2+\tilde{p}^2)$. As $\gamma$ is increased from zero,
the current increases to a maximum and then falls off, which has the
form of stochastic resonance~\cite{allison01b}. The position of this
extremum can be obtained from the condition, $\partial I_\gamma
/\partial \gamma = 0$. $\gamma$ varies a little from $0.408$ for
$p=0.1$ to $0.423$ for $p=0.5$.

\begin{figure}
  \resizebox{84.67mm}{!}{\includegraphics*{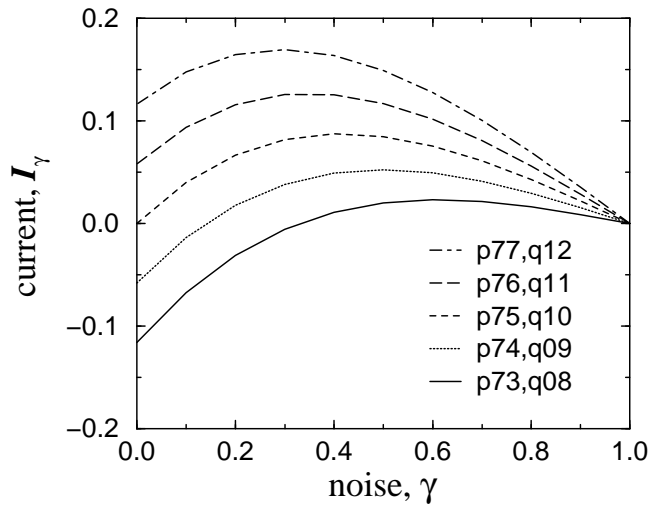}}
  \caption
  {The probability current $I_\gamma$, versus noise parameter,
  $\gamma$, off the null-surface. The bottom curve corresponds to
  $p=0.73$ and $q=0.08$ and the top to $p=0.77$ and $q=0.12$.}
  \label{fig:fig4}
\end{figure}

Fig.~\ref{fig:fig4} shows the net current, $I_\gamma$, versus noise $\gamma$
when the system is not on the null-surface any more.  In the off-balance
region, $q \ne \tilde{p}^2 / (p^2 + \tilde{p}^2)$, the net current is not
zero for $\gamma=0$ but still should be zero for $\gamma=1$ and the
intermediate behavior is qualitatively the same as the balanced behavior. The
actual values of $p$ and $q$ for the various curves in Fig.~\ref{fig:fig4}
are in linear increments of $0.01$ for $p$ and $q$. The top curve has
parameters $p=0.77$ and $q=0.12$. The bottom curve has parameters $p=0.73$
and $q=0.08$.

We generalize the MBR by introducing a bias into the added noise. The walker
takes a right step with probability of $0.5-\epsilon$ and a left step with
probability of $0.5+\epsilon$. For $\epsilon \ne 0$ this noise introduces
non-zero net current. The new parameter, $\epsilon$, is essentially a measure
of the bias in the added noise.



We can generalize this model to a system of size $N $ by repeating the
unit cell of modulo-3 $N$ times with a periodic boundary condition. In
this case, the periodic potential ensures
$ p_{k}(t)=p_{k+3n}(t)~\forall~n=0,\pm1,\pm2, \cdots.   $
Because of the normalization condition, $\sum_{k=1}^{N} P_k(t) =1$,
the current will be reduced by factor of $N$. Otherwise, the
corresponding master equations and solutions are exactly same as the
minimal model.

For different moduli, in principle, we can also set up the master
equations and solve them exactly by matrix inversion for the set of
linear equations. It can be shown that these results have
qualitatively the same statistical behavior as the 3-state MBR. Note
that for even number moduli there are oscillatory non-stationary
solutions.

The transformation in Eq.~\ref{eq:parameter-change} tells us
effectively that the MBR gives the same results as a biased random walk, where
the transition probability is not symmetric but biased. Although this
analogy can be used to investigate the characteristics of MBR, it is
absolutely impossible to determine whether the system is itself a biased
random walk or an MBR by analyzing the result of measurements, without prior
knowledge that the model is a combination of a balanced unbiased random walk
and added random external noise. This makes the MBR valuable for
understanding the minimal features of the discrete-time Brownian ratchet.
The MBR has applicability in discrete-time processes where
the transition probabilities do not fluctuate in time, such as in game and
computation theory where transitions occur at precisely defined times.

Funding from GTECH Australasia, Sir Ross and Sir Keith Smith Fund and
the Australian Research Council (ARC) is gratefully acknowledged.

\end{document}